# Gate-controllable electronic trap detection in dielectrics

Sandip Mondal[1,*] Tathagata Paul[1], Arindam Ghosh[1] and V. Venkataraman[1]

*Abstract*— Gate controllable electronic trap detection method has been demonstrated by regulating the gate potential of MIS devices. This method is based on shift of capacitance–voltage (CV) curve as well as flatband voltage ($V_{FB}$) measure in <10 µs due to injection or ejection of electrons through the metal gate. Using this method, an electronic trap energy distribution was measured in the $HfO_2$ dielectric film and it confirms a maximum number of traps ($\Delta N_T$) of $1.7 \times 10^{12}$ cm$^{-2}$ corresponding to an energy level ($\Delta E_{IL}$) of 0.45 eV above silicon conduction band (Si-$E_{CB}$). In comparison, $ZrO_2$-based MIS devices showed a broader distribution of electronic traps throughout the band gap. However, $HfO_2$ contained more than 60% traps in deep level compared to 50% in $ZrO_2$, which establishes the effects of material variation.

## I. INTRODUCTION

It is desirable to operate floating gate flash memory devices using the gate controllable method that increases reliability and endurance[1], [2],[3] and reduces inter-cell interference as the devices are scaled down[4]–[9]. The performance of gate controllable memory devices depends on the number of traps available in the floating gate, which is mostly fabricated with dielectric materials that contain a large number of electronic traps[10]. Hence, it is important to characterize the traps in dielectrics before fabricating flash memory devices. Several methods to detect traps in dielectrics have been proposed[11]–[13]. However, most of these methods are based on channel injection mechanism. The channel injection mechanism[14], [15] is useful but requires devices with FET architecture, where the floating gate dielectrics are located on an additional $SiO_2$ layer.

In this work, we present a gate-controllable (GC) charge injection and ejection method in combination with high-speed capacitance–voltage measurement to locate the traps in dielectric films by using two-terminal MIS structure. The high-speed-CV measurement was performed after every erase/program (E/P) operation to detect the trap distribution due to a shift in $V_{FB}$. Moreover, a major advantage of this method is that it is independent of the band bending in silicon, thus eliminating the contribution due to substrate or interface defects. In addition, the GC method is capable of measuring the trapped charges throughout the bulk dielectric by regulating the E/P voltage or injection time. Using this method, many CV curves were measured and the shift in $V_{FB}$ for different E/P pulses was analyzed; this extracted the distribution of traps in wide band gap dielectrics such as $HfO_2$ as well as $ZrO_2$ and compared, which identifies the effects of material variation.

## II. DEVICE FABRICATION

Fig 1(a) demonstrates the structure of gate-controllable MIS where Al-metal and p-type silicon (4.8 Ω.cm$^{-1}$) were used as top electrode and bottom semiconductor respectively. The MIS devices were fabricated using two ($HfO_2$[16] and $ZrO_2$[17]) different types of dielectrics followed by sol-gel processed technique. The spin-coated film was heated immediately at 200 °C for 1 min. The typical thickness of the dielectric films was 30 nm. All films were subjected to 900 °C for 1 h. The top electrode contact was fabricated by thermally depositing 200 nm thick Al-metal through a shadow mask. The large area ohmic back contact was formed by using silver paint for the purpose of electrical measurement.

## III. RESULTS AND DISCUSSION

CV measurement was performed on an MIS device ($HfO_2$) at different frequency (1–100 kHz) domains by using the LCR (HIOKI 3532) meter (Fig. 1b). The device showed a large hysteresis in CV curve with a $\Delta V_{FB}$ of 1.8 V for a gate sweep voltage of ± 5 V, which corresponds to an electronic trap density of $1.9 \times 10^{12}$ cm$^{-2}$. The $\Delta V_{FB}$ is independent of the measurement frequency. Moreover, the device showed a donor-like trap density since an opposite $V_{FB}$ shift was observed with respect to polarity of the gate bias. For example, a positive change in $V_{FB}$ was obtained when the gate voltage was swept from -5 V to +5 V and vice versa. These devices are gate-controlled, where the charge injection or trap occupancy is determined by Fermi level of the metal gate. Such devices are highly useful for memory device applications due their gate controlling nature.

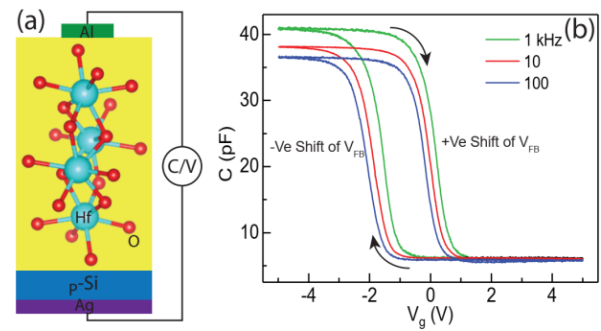

Fig 1: (a) Schematic structure of MIS device with $HfO_2$ film on silicon substrate (p-Si). The molecular structure is created by using the VESTA (https://jp-minerals.org/). (b) The CV curve was measured at different frequency domain by sweeping the gate potential from -5 V to +5 V and back.

Sandip Mondal, Tathagata Paul, Arindam Ghosh and V. Venkataraman is with Dept. of Physics, Indian Institute of Science, Bangalore 560012, India.
Sandip Mondal is currently with School of Materials Engineering, Purdue University, West Lafayette, IN 47907, USA (e-mail: sandip@iisc.ac.in)

The electronic band diagram[18], [19] of the MIS structure that was used to characterize electronic trap states in wide band gap dielectrics is illustrated in Fig. 2(a). The electron affinity of the metal, dielectric and semiconductor are defined as $\phi_M$, $\chi$ and $\phi_{Si}$ respectively. The Fermi level difference between the metal and silicon is $E_F$ (Al–Si) = 0.7 eV. The grey-shaded region demonstrates the completely-filled trap state upto the metal Fermi level ($E_F$ of Al). Some of these traps are inaccessible; hence, it is difficult to remove electrons due to their location at deeper levels of the energy band. Initially, a high positive gate voltage (Vg >> 0) was applied for an extended time period to release electrons from the trap states through the metal Fermi level upto the silicon Fermi level as shown in Fig 2(b). The triangular yellow-shaded region indicates the accessible traps in the dielectric that can be unfilled. The electronic traps that are located below the yellow region (grey) are called deep level or inaccessible traps.

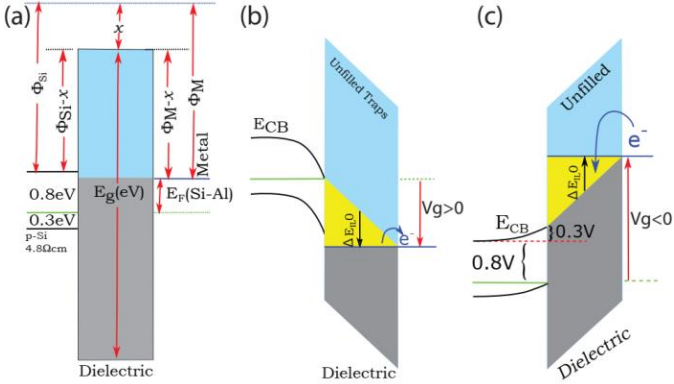

Fig. 2. Energy band diagram of the MIS devices at different biasing condition. (a) Illustration of flatband condition before applying any potential. (b) During discharge of electronic traps a positive gate voltage (V$g$ > 0) is applied, where, the yellow shaded region indicating the accessible traps in the dielectric. (c) The band bending during injection of electron to trap states due to negative gate bias (V$g$ < 0) for a longer time.

First the band diagram with a trap at location $\Delta E_{IL}$ (eV) from the Si conduction band was drawn for flat band condition as similar to method was adopted for the shift of threshold voltage ($V_{th}$) of TFT [12]. When a program voltage is applied (longer time), the band diagram was redrawn with the average trap location shifted in energy. By matching the shifted trap location with the Fermi level of the metal, the original energy value $\Delta E_{IL}$ (eV) could be calculated. This was repeated for every program voltage. The mean positions of the accessible traps ($E_{IL}$) are measured from the silicon conduction band edge (Si-$E_{CB}$). For example, during the application of a large +Ve gate voltage (Vg >> 0), all the accessible traps become empty due to removal of electron from the dielectric layer; hence, the mean position of the trap state is equivalent to $\Delta E_{IL}$ = $V_g/2$. However, the accessible traps that are located above the Si-$E_{CB}$ can be filled through injecting electrons by applying a negative gate voltage (injection voltage). The band bending of MIS changes according to amplitude of the injection voltage as shown in Fig. 2(c). In this scenario, the measured $\Delta E_{IL} = [V_g-(Si-E_{BB})-(Si-E_{CB-F})]/2$, where Si-$E_{BB}$ is the silicon band bending of 0.3 eV and Si-$E_{CB-F}$ indicates the difference between the silicon Fermi level and conduction band (~ 0.8V). Using the gate-controlling method, the electronic trap energy density spectrum was measured in the wide band gap dielectric films.

The GC method was verified using sequential emission and injection charges through the gate electrode (Fig 3a). Initially, a positive gate voltage (Inset: +$V_1$) was applied to the MIS device to remove electrons from the trap state. Hence, the CV curve and corresponding $V_{FB}$ were moved towards the negative direction in the voltage axis. This confirms the removal of electrons from the trap states. Following this, an injection voltage (-$V_1$) of -10 V was applied to fill the trap states. The CV curve was measured using a triangular voltage pulse[20] (< 10 μs) after every injection pulse (Inset of Fig 3a). The device displayed a chronological shift in $V_{FB}$ after each application of -$V_1$ (Fig 3b). Within 10 s, a large number of traps ($\Delta N_T$) got filled by electrons and hence, the change in $V_{FB}$ turned towards saturation. This implies the availability of a large number of gate-controllable traps in HfO$_2$ dielectrics. The $\Delta N_T$ in a particular trap energy state ($\Delta E_{IL}$) is calculated by using, $\Delta N_T = (C_{ox} \times \Delta V_{FB})/qA$[21], where $C_{ox}$ is the accumulation capacitance dielectric, $\Delta V_{FB}$ is the difference in flatband voltage between two states, q is the electronic charge and A is area of the device.

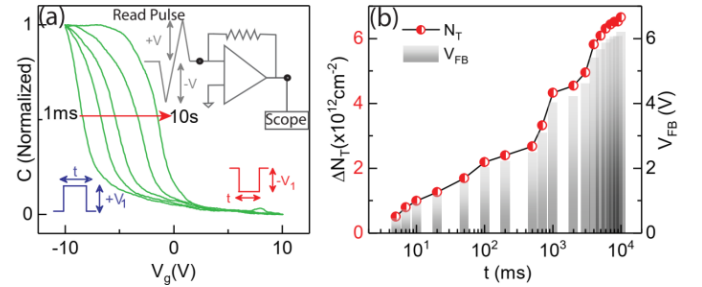

Fig. 3. (a) The CV curve is measured on HfO$_2$ based MIS devices after application of every injection voltage pulse (-$V_1$ = -10 V) till 10 s. Inset: The CV curve was measured by applying the triangular voltage pulses. (b) Change in $V_{FB}$ and $\Delta N_T$ due to application of –$V_1$ at different time (t = 1-10$^4$ ms).

A sequential decrease in $V_1$ was applied to the MIS device for longer time duration (t = 10$^4$ s) to detect the different level electronic trap states (Fig 4a). Simultaneously, the CV curve was measured at every injection point. Before applying $V_1$, the accessible traps were completely unfilled by applying +10 V for 10$^4$ s. A complete unfilled depth of traps level is confirmed once the shift of flatband voltage ($\Delta V_{FB}$) is stopped in-spite of applying higher $V_1$ for very longer time. For example, the $V_{FB}$ for final three consecutive points at t = 10$^2$, 10$^3$ and 10$^4$ s are same as 11.36 V when the charging voltage is 6 V. Then, the injection process was initiated by applying $V_1$. It was observed that $V_{FB}$ shifts towards the positive voltage with respect to negative of $V_1$ and vice versa. Such a $V_{FB}$ shift also confirms the gate-controllable mechanism of the device to detect electronic traps. The electronic traps usually become saturated when the injection time period is long enough (t > 100 s). For example, $\Delta V_{FB}$ for the last three successive data points at t =

8000, 9000 and 10000 s were similar to -3.1 V, when the applied injection voltage was $V_1$ = -4 V. A bar plot is shown in Fig 4b, which compares discrete categories of $\Delta V_{FB}$ at three different time. Such $\Delta V_{FB}$ is collected when $V_1$ is applied for longer time duration. Moreover, the traps normally get saturated within 100 s; hence, the $\Delta V_{FB}$ is similar even after application of $V_1$ for $10^4$ s. The $\Delta V_{FB}$ keeps increasing by lowering the $V_1$. The highest $\Delta V_{FB}$ of 1.8 V was obtained when a $V_1$ = -2 V was applied for $10^4$ s.

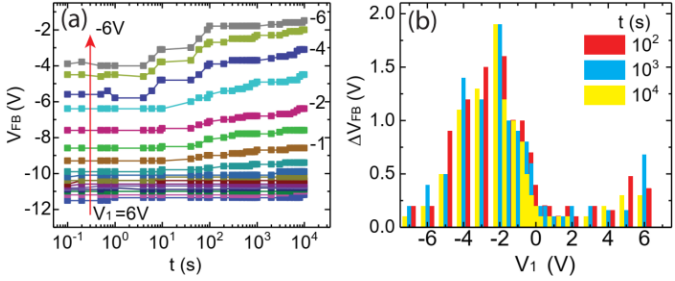

Fig. 4. (a) Different flatband position due to injection of charge to $HfO_2$ dielectric by applying the injection voltage ($V_1$) from 6 V to -6 V for $10^4$ s. (b) Shift $\Delta V_{FB}$ with respect to $V_1$ measured for different injection time.

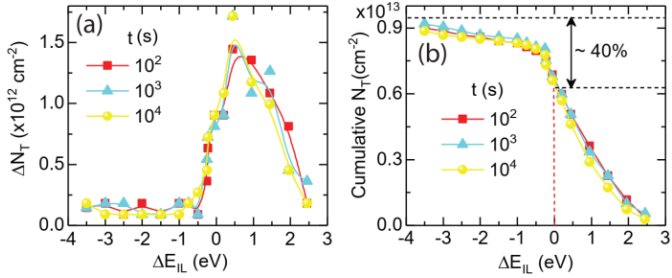

Fig. 5. (a) Spectral energy distribution of electronic trap state compared for different injection time for $HfO_2$ dielectric. (b) A cumulative trap density is extracted with respect to energy level of traps for the different injection time.

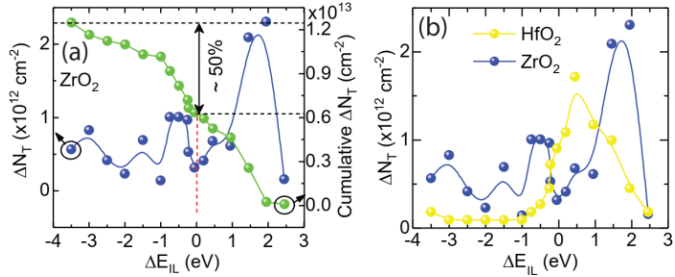

Fig. 6. (a) Distribution of traps energy and cumulative trap density at different energy levels for $ZrO_2$ dielectric. 50 % of traps are located above the Si-$E_{CB}$. (b) Comparison of trap energy spectrum for $HfO_2$ and $ZrO_2$ dielectric.

It was observed that the number of electronic traps increases with respect to the negative injection voltage (< $V_1$). A maximum number of electronic traps ($\Delta N_T$ = $1.7 \times 10^{12}$ cm$^{-2}$) in the $HfO_2$ dielectric film are accessible with $V_1$ = -2 V, which corresponds to a $\Delta E_{IL}$ at 0.45 eV[12] that is above Si-$E_{CB}$. The $\Delta N_T$ reduces with even decrease in $V_1$. Hence, entire electronic trap energy spectrum was measured by using the gate-controlling method. An accumulative number of traps were calculated with respect to $\Delta E_{IL}$ (Fig. 5b). It was observed that only 40 % traps were available above Si-$E_{CB}$, which is also referred to as shallow level traps. However, a large number of traps (~ 60%) were available below Si-$E_{CB}$, also called as deep level traps in the dielectric that are highly useful for flash memory applications. Hence, all shallow and deep level traps were measured in the $HfO_2$ dielectric materials by using the GC method.

TABLE I
ELECTRONIC TRAP MEASUREMENT METHOD COMPARISON

| Electronic Trap Measurement Method | Trap Energy [$ZrO_2$ & $HfO_2$]$^\Omega$ | Number of Traps [$ZrO_2$ & $HfO_2$]$^E$ | Ref. |
|---|---|---|---|
| C-f Modeling | 0.1 & 0.1 | 1 & 4 | [22] |
| $g_m$-f | # & # | # & 3 | [23] |
| Photoluminescence | # & 1.1 | # & < 4 | [24] |
| Multilayer | 0.6 & 0.70 | .04 & 1.3 | [25] |
| DMP technique | # & 0.45 | # & 2 | [12] |
| Gate-Controllable | 1.95 & 0.45 | 2.3 & 1.7 | * |

$\Omega$ = Unit: eV; E = Unit: ($\times 10^{12}$) cm$^{-2}$ eV$^{-1}$; # = Data is not available; * = Present work.

An electronic trap energy spectrum was also measured in the $ZrO_2$ dielectric film by using the gate-controllable electronic trap detection method (Fig. 6a). This dielectric demonstrated a wide distribution of electronic trap energy ($\Delta E_{IL}$) of -3 eV, -1.5 eV, -0.5 eV, 0.45 eV and 1.95 eV corresponding to the traps density $0.8 \times 10^{12}$, $0.7 \times 10^{12}$, $1 \times 10^{12}$, $0.7 \times 10^{12}$ and $2.3 \times 10^{12}$ cm$^{-2}$ respectively. Initial three peaks were above and remaining two peaks are located below Si-$E_{CB}$. The accumulative electronic trap distribution confirmed that nearly 50 % electronic traps are situated above Si-$E_{CB}$. The maximum numbers of electronic traps were located at energy of 1.95 eV, which is much higher than the $HfO_2$ dielectric (Fig. 6b). Also, the cumulative electronic trap density is slightly higher in $ZrO_2$ than the $HfO_2$ due to availability of higher number electronic traps in energy spectrum (peak at 0.45 eV for $HfO_2$ and 1.95 eV for $ZrO_2$). The obtain result from gate-controlling trap detection method is also compared with other detection method (TABLE I), which shows these traps characteristics are in good agreement with previous report on same dielectric[12]. However, the number of deep level traps in $ZrO_2$ was lower than the $HfO_2$ dielectric film. Thus, using the gate-controlling electronic trap detection method, the electronic trap energy spectrum of $HfO_2$ and $ZrO_2$ dielectrics was measured and compared.

## IV. CONCLUSION

A highly sensitive gate controllable trap detection method is demonstrated to detect shallow and deep level electronic traps in dielectrics. The method is simple, independent of substrate material and captures the trap energy signature in dielectrics; this can assist in material selection during technology development. Using this method, trap energy distributions in $HfO_2$ and $ZrO_2$ dielectrics were extracted and compared. The maximum number of traps was detected at an energy level of 0.45 eV, which is above the Si-$E_{CB}$ for $HfO_2$ dielectrics. On the other hand, $ZrO_2$ dielectric comprises of wide distribution of electronic traps. Moreover, $HfO_2$ shows a large number of deeper level traps compared to $ZrO_2$. Such an electronic trap energy spectrum is highly useful for characterizing the performance of charge-trapping memory devices.